\documentstyle[psfig]{l-aa}
\begin{document}

   \thesaurus{03(11.05.1, 11.11.1, 11.01.1, 11.06.1, 
   11.09.1 NGC~4816, 11.09.1 IC~4051)} 

\title{The kinematically peculiar cores of the Coma cluster early -- type 
galaxies NGC 4816 and IC 4051}

\author{D. Mehlert\inst{1}\fnmsep\thanks{Visiting astronomer at the
German-Spanish Astronomical Center, Calar Alto, operated by the
Max-Planck-Institut f\"ur Astronomie, Heidelberg jointly with the Spanish
National Commission for Astronomy.}\fnmsep\thanks{Visiting astronomer
at the Michigan - Dartmouth - M.I.T. Observatory, Kitt Peak, Arizona, operated by
a consortium of the University of Michigan, Dartmouth College and the
Massachusetts Institute of Technology.}, R.P. Saglia\inst{1}, R. 
Bender\inst{1}, G. Wegner\inst{2}
          }

   \offprints{D. Mehlert (email: mehlert@usm.uni-muenchen.de)}

   \institute{{Universit\"atssternwarte M\"unchen, D-81679 M\"unchen,
             Germany}
        \and{Department of Physics and Astronomy, 6127 Wilder Laboratory, 
	     Dartmouth College, Hanover, NH 03755-3528, USA}}

\date{accepted to be published in A\&A, main journal}

   \maketitle \markboth{Mehlert et al.: 
   The kinematically peculiar cores of NGC 4816 and IC 4051}{}
   
 \begin{abstract}

The Coma cluster is one of the richest known cluster of galaxies, spanning
about 4 dex in density. Hence it is the ideal place to study the structure
of galaxies as a function of environmental density in order to constrain the
theories of galaxy formation and evolution.  For a magnitude limited
sample of 35 E and S0 galaxies we obtained long slit spectra to derive the 
rotation curves, the velocity dispersion profiles and the radial gradients of 
the Mg, Fe and H$\beta$ line indices.
Here we report on two early -- type
galaxies which turned out to host the largest kinematically
peculiar cores yet found in ''normal'' early -- type galaxies:
NGC~4816 hosts a decoupled counter rotating
core with a radial extension along
the major axis of 2.7~kpc, while IC~4051 has a
co-rotating peculiar core with a sizes of 3.4~kpc.
We combine our data with HST
photometry and show that both cores are flattened central stellar
disks which
contribute less than 1~\% to the total V band light of the
galaxies, but are nevertheless 
conspicuous (1 -- 2 $\times$ 10$^9$ L$_{\odot}$).
The metallicity of the cores is 0.25 dex super solar
and drops to solar and sub solar in the outer part of NGC~4816 and
IC~4051, respectively. The mean stellar population in both central disks is
old (8~--~12~Gyr) and highly overabundant in Mg$_b$
relative to $<$Fe$>$ ($\approx$~0.5~dex). We discuss
the evidence that these central disks formed via dissipational major
merger events.

\keywords{galaxies: elliptical and lenticular, cD --   
kinematics and dynamics -- abundances -- formations -- 
individual: NGC 4816 -- IC4051}
   \end{abstract}

\section{Introduction}

Kinematically peculiar cores are generally understood as being fossil
fingerprints of the merging - and thus formation - history of the
host galaxies (Kormendy 1984, Franx \& Illingworth 1988, Bender 1988,
Jedrzejewski \& Schechter 1988). They have mostly been explained
by either accretion of
small spheroidal satellites by larger ellipticals (Kormendy 1984,
Balcells \& Quinn 1990, Balcells 1991) or as the result of major
merging events by star dominated systems
(Schweizer 1990, Barnes 1992a, Bender \& Surma 1992, Hernquist 1993,
Heyl et~al. 1994).  Up to
now kinematically decoupled cores with radial extensions between 0.2 -
2.7~kpc have been found in a number of elliptical galaxies - lying in
the field and in groups as well as in clusters (Bender 1990; see
Sect.~5 for a detailed list).\\
Most peculiar cores have been shown to be due to a rapidly spinning
disk -- like component which presumably formed dissipatively (Bender 1988,
1990, Franx \& Illingworth 1988, Rix \& White 1992, Surma \& Bender 1995).
As shown by Bender \& Surma (1992) the high absorption line
strengths of the decoupled regions further argue against  accretion of
compact small galaxies and suggest a scenario analogous to massive
spiral -- spiral mergers (see also Davies et al. 1993). In this model the
pre-enriched gas in the two merging components lost its angular
momentum, settled in the central region and underwent a strong star
formation phase. Though spiral -- spiral mergers may still form
ellipticals today (Kormendy \& Sanders 1992), it is unlikely that
most decoupled cores formed recently. Their location in the
Mg -- $\sigma$ diagram indicates that the corresponding ellipticals 
have ages similar to normal ellipticals.
Recent HST images of kinematically decoupled cores (Carollo et al. 1997)
show that 45 - 60 \% of them are in fact stellar central disks and
confirm the idea, that most of them indeed have formed by dissipational
merging processes.\\
Finally HST imaging by Jaffe et al. (1994), Lauer et al. (1995) \& 
Faber et al. (1997) has shown that the most luminous,
anisotropic, boxy galaxies possess surface profiles with shallow cores 
while low luminous, rotationally flattened, disky ellipticals have 
power-law profiles (confirming earlier claims by Nieto et al. 1991). 
\\ 

Here we present
kinematical and line index profiles as well as HST surface brightness
profiles for the two Coma galaxies IC 4051 and NGC 4816, which host
the largest cores yet found in early -- type galaxies.\\ In Sect.~2
we describe the spectroscopic observations, data reductions and
the results of the kinematical analysis and line strength
measurements.  The photometric data obtained with the Hubble Space
Telescope (HST) and the MDM 1.3~m telescope are described in 
Sect.~3, where we derive the surface brightness profiles and other
photometric parameters. 
We summarize the combined spectroscopic and photometric results in 
Sect.~4. Finally we discuss formation scenarios for NGC 4816 and IC 4051
in Sect.~5.

\section{Spectroscopy}

\subsection{Observations \& data analysis}

{\small
\begin{table*}
\begin{center}
\begin{tabular}{|c|c|c|c|c|c|c|c|c|c|}
\hline
Galaxy & Telescope & Detector & $\lambda\lambda$ -- range  & Scaling &
Slit width & Resolution & Exp. time & PA & seeing  \\
 & & & [\AA] & [''/pix] &  [''] & & [s] & (N to E) & (FWHM) \\ 
\hline 
\hline
{\bf NGC 4816} & CA & TI & 4730 -- 5700 & 0.89  & 3.6 & 1.17 \AA & 7000 &
78$^{\circ}$ & 1.2''\\
 & 3.5 m & 1024 x 1024 &   &  & & 67.6 km/s &  &  & \\
\hline
{\bf IC 4051 }& MDM & TI & 4300 -- 6540 & 0.77  & 1.7  & 2.3 \AA & 10800 &
106$^{\circ}$ & 1.8''\\
 & 2.4 m & 1024 x1024 &   & &  & 149.7 km/s &  & & \\
\hline
\end{tabular}
\caption{Setup of the observing runs} \label{tobs}
\end{center}
\end{table*}
}
Long-slit spectra centered on the 5170 \AA\ Mg triplet of NGC~4816 and
IC 4051 were taken along their major axes. 
NGC~4816 was observed at the 3.5~m telescope of the
German-Spanish Astronomical Center on Calar Alto (CA) in 1996 (see
Table~\ref{tobs}). IC~4051 was observed with
the 2.4~m telescope of the Michigan-Dartmouth-M.I.T. (MDM) observatory on
Kitt Peak in 1995. Additionally, during both runs we obtained
spectra for ten K -- Giant template stars, trailed and wiggled across the slit.
The standard reduction (bias \& dark subtraction,
flat fielding, wavelength calibration, sky subtraction, correction for
CCD misalignment) was carried out with the image processing package
MIDAS provided by ESO.\\ After continuum subtraction and radial
binning along the slit we achieved a S/N $\geq$ 40/\AA\ for all of our data. We
determined the line-of-sight-velocity-distributions (LOSVDs) by using
the Fourier Correlation Quotient (FCQ) method (Bender 1990), which
provides us with the stellar rotational velocities v$_{rot}$, the
velocity dispersions $\sigma$ and the first orders of asymmetric
(H$_3$) and symmetric (H$_4$) deviations of the LOSVDs from real
Gaussian profiles (van der Marel \& Franx 1993 ; Bender et al. 1994). 
As expected, we find that the FCQ method is little
influenced by template mismatching (Bender 1990). 
The ``best'' template star minimizing the asymmetric
deviations (H$_3$) of the LOSDV from a real Gaussian profiles turned out to
be HR 6817.
Following Bender et al. (1994), Monte Carlo simulations
were made to find the best fit-order for the continuum, to check for
systematic effects and to estimate errors. For NGC 4816 two exposures with an
atmospherical seeing of FWHM = 1.2'' and one with FWHM = 3.7'' were available.
The latter one was excluded for the final analysis, but was used to check,
whether the large slit width (3.6'') affected the spectral resolution 
significantly for good seeing conditions. The three individual
exposures did
not produce significant differences within the errors.\\
Line strength indices were
derived following Faber et al. (1985) and Worthey (1992) from flux
calibrated spectra, rebinned radially as before. Furthermore, we
corrected for velocity dispersion broadening 
and calibrated our measurements to
the Lick system using stars from Faber et al. (1985). The
errors are derived from photon statistics and CCD read-out noise.

\begin{figure}  
\psfig{figure=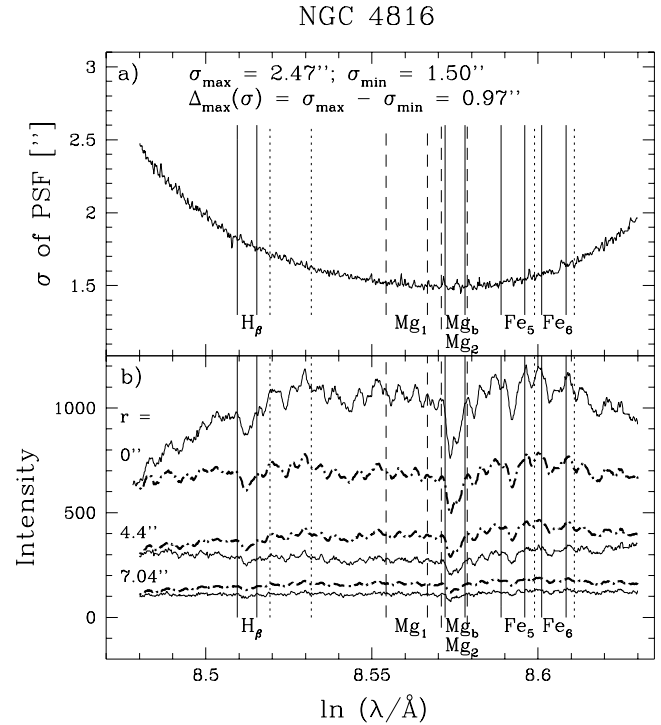,height=10cm}
\caption{\label{focus}
{\bf (a)} The measured Gauss width $\sigma$ at each wavelength for one single
frame of NGC 4816. We measured $\sigma$ inside a radius of 7'', which
turned out to be the radius at which the focus variation starts to effect the 
continuum of the measured spectra.
For H$_{\beta}$, Mg$_b$, Fe$_5$ and Fe$_6$ the line 
windows defined by Faber et al. (1985) are indicated (solid lines). For
the Mg$_1$ and Mg$_2$ indices the line windows (long dashed lines) 
and the windows for their pseudo continuum (short dashed lines) are shown.
{\bf (b)} The spectra of NGC 4816 at different radii before (thin solid line)
and after (fat dashed dotted line) the focus correction. The different
continua shapes in the uncorrected spectra are evident. Like in Fig.~{\bf (a)}
the windows for the measured indices are indicated.}
\end{figure}

\noindent The line index 
profiles for NGC 4816 needed further treatment. At small radii some of them
show unreal features caused
by the varying focus of the spectrograph in the dispersion direction,
which depends slightly on the
wavelength.  Following Gonzales (1993), this effect is only
detectable if the variation of the focus' point spread function (PSF)
is dominant compared to
the atmospherical seeing (FWHM).
For NGC 4816 the focus variation was the dominant effect
and we developed a new method to correct for this ``focus
effect'': For each single exposure of NGC 4816 we
measured the central Gaussian width $\sigma(\lambda)$ of the radial profile at each
wavelength (Fig.~\ref{focus}a) and noted that $\sigma$ is
larger at the blue and red end of the spectra compared to the central
wavelength. Therefore
blue and red light is spread more radially than light in the
center of the spectrum. Consequently, the spectrum's continuum shape
varies with radius (see Fig.~\ref{focus}b; thin solid line) and leads to
unreal central features in the measured line index profiles.  The
Mg$_1$ and Mg$_2$ indices defined by Faber et al. (1985) are affected
most, because their pseudo continua are $\approx$ 200 \AA\ apart from
the line windows.  For all the other line indices the continuum and
line windows are close to each other. H$_\beta$ and the Fe indices are
in the red and blue wing of the spectra and thus slightly affected,
while the Mg$_b$ index at the central wavelength is almost
unaffected. To correct for this focus
variation in the exposures of NGC 4816
we convolved all radial profiles with the wavelength depending
width $\sigma_{broad} = (\sigma_{max}^2 - \sigma^2(\lambda))^{1/2}$ 
(see Fig.~\ref{focus}a).
As a result the radial profiles have the
width $\sigma_{max}$ = 2.47'' at all wavelength.
Furthermore, the continuum shapes of the spectra are the same for all
radii (see  Fig.~\ref{focus}b; fat dashed dotted line) and the artificial
features in the line index gradients disappear
(see Mehlert 1998 for a more detailed description).
For the spectra of NGC 4816 the maximum broadening width is
$\sigma^{broad}_{max} = (\sigma_{max}^2 - \sigma_{min}^2)^{1/2} = $1.96''
($\sigma_{min} = $1.5''; see Fig.~\ref{focus}a).
The resulting PSF for the measured line index profiles (shown in
Fig.~\ref{spec4816}b) has
$\sigma_{res} = ((\sigma^{broad}_{max})^2 + \sigma_{seeing}^2)^{1/2} = 2.0$''
and FWHM$_{res}$ = 4.7'' .\\
For IC 4051 no ``focus correction'' was necessary,
because the atmospherical seeing was dominant during the
observations.\\

\subsection{Kinematical profiles and line index gradients}

\begin{figure*}  
\begin{center} \parbox{14cm}{ 
\psfig{figure=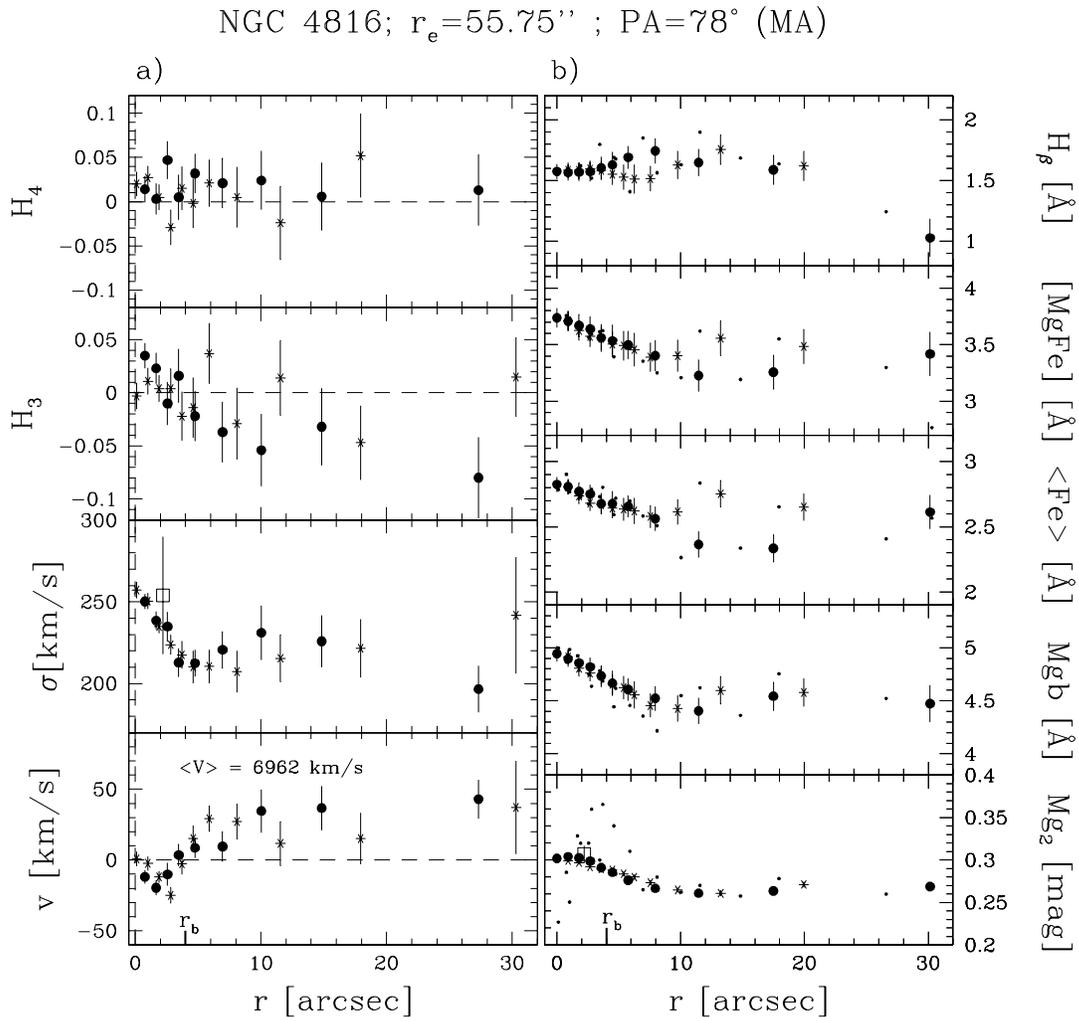,width=14cm}
\caption{\label{spec4816}
{\bf (a)} Kinematical parameters and measured line indices {\bf (b)}
for NGC 4816 along the major axis. The curves are folded around the nucleus of
the galaxy and different symbols refer to different sides.
For {\bf (b)} the profiles before (small dots) and after
focus correction (large symbols) are shown.
Open squares are results from
Davies et al. (1987) who used a 4''$\times$4'' aperture. We determined the
equivalent radius along our used slit via r$_{d87}$ = 0.5 $\times$
(4''$\times$4''/ slitwidth).
Panels from bottom to top: {\bf (a)} (1) Rotation velocity v(r) relative to the system velocity,
(2) velocity dispersion $\sigma$(r), (3-4) Gauss-Hermite
moments H$_3$ and H$_4$.
{\bf (b)} (1-2) Mg$_2$ and Mg$_b$, (3) mean Fe index $<$Fe$>$ = (Fe$_5$ + Fe$_6$)/2,
(4) the combined index $[$MgFe$]$ = (Mg$_b<$Fe$>$)$^{1/2}$, (5) H$_{\beta}$ index. 
}}
\end{center}
\end{figure*}

\begin{figure*}  
\begin{center} \parbox{14cm}{ 
\psfig{figure=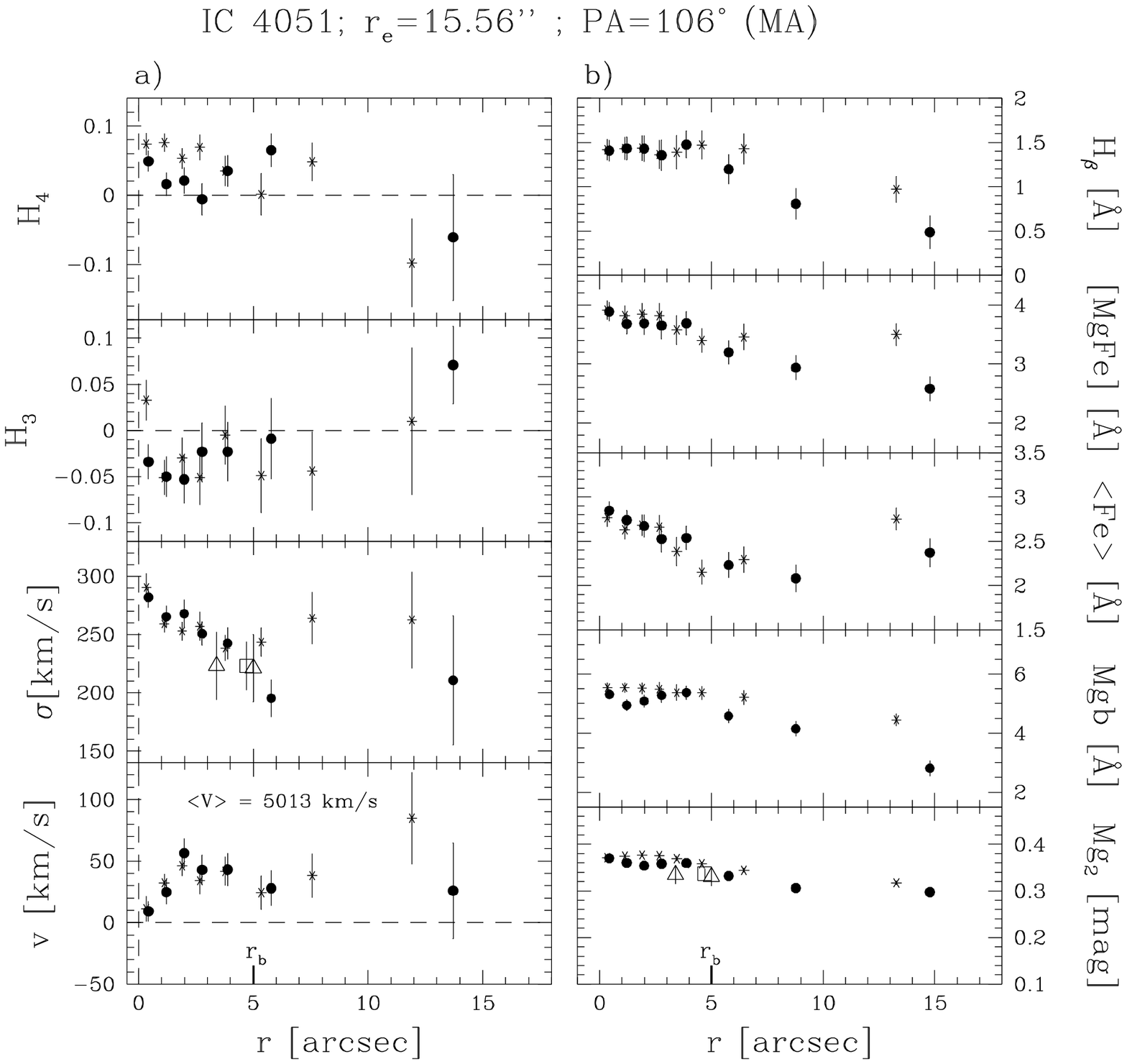,width=14cm}
\caption{\label{spec4051}
Kinematical parameters {\bf (a)} and measured line indices {\bf (b)}
for IC 4051 along the major axis. The curves are folded around the nucleus
of the
galaxy with different symbols referring to different sides. Open triangles
represent data from J{\o}rgensen et al. (1995) and open squares those from
Davies et al. (1987). The latter used a 4''$\times$4'' aperture. We determined the
equivalent radius along our used slit via r$_{d87}$ = 0.5 $\times$
(4''$\times$4''/ slitwidth).
Panels from bottom to top: {\bf (a)} (1) Rotation velocity v(r) relative to
the system velocity,
(2) velocity dispersion $\sigma$(r), (3-4) Gauss-Hermite
moments H$_3$ and H$_4$.
{\bf (b)} (1-2) Mg$_2$ and Mg$_b$, (3) mean Fe index $<$Fe$>$ = (Fe$_5$ + Fe$_6$)/2,
(4) the combined index $[$MgFe$]$ = (Mg$_b<$Fe$>$)$^{1/2}$, H$_{\beta}$ index. 
}}
\end{center}
\end{figure*}

Figs. \ref{spec4816} and \ref{spec4051} show the kinematical profiles
(a) together with the line indices (b) of NGC 4816 and IC 4051,
respectively.\\

\noindent{\bf NGC 4816:}

\noindent This galaxy contains a kinematically decoupled core, which
is counter-rotating with respect to
the main spheroid at a maximum velocity v$_{max}$ = 30 km/s. The kinematical
break radius is r$_b$~=~4'', which is 2.7 kpc in the Coma cluster
(1'~=~40 kpc; H$_o$~=~50 km/s/Mpc).
The velocity dispersion is $\sigma_{b}$~=~220~km/s
at r$_b$ and its gradient gets much
steeper inside the core,
rising to a central value of $\sigma_o \approx$ 260 km/s.
Finally, H$_3$ changes sign at the core
radius.  The central velocity dispersion measured by Davies et
al. (1987) lies a bit above our profile, but still agrees with it
within the errors.

For NGC 4816 the line indices measured with (large symbols) and
without (small dots) the focus correction described in Sect.~2 are
shown and the need and improvement for the Mg$_2$ index is obvious.
Without focus correction the profile of the Mg$_2$ index is
disturbed out to 7''.\\
For the focus corrected profiles we find an increase of slope in the Mg$_2$
as well as in the Mg$_b$ line
profile inside 8'', which is larger than r$_b$ (see section~\ref{results} 
for explanation).
Beyond 8'' the Mg$_b$ index is constant.  The mean iron
index $<$Fe$>$ = (Fe$_5$ + Fe$_6$) / 2 and the combined index  $[$MgFe$]$ =
(Mg$_b<$Fe$>$)$^{1/2}$ do also increase inside
8'' and stay constant beyond that point. The H$_{\beta}$ index shows a
slight increase towards the outer parts of the galaxy, but is dropping
at the very outer (less reliable) data point.  The Mg$_2$ value
derived by Davies et al. (1987) is a little larger than our data, but
still agrees within the errors.

\noindent{\bf IC 4051:}

\noindent This galaxy contains a peculiar core which is co-rotating
with the main spheroid at a
maximum projected velocity v$_{max}$~=~60~km/s. The core's kinematics
are detectable to a radius of 5 '', which is 3.4 kpc in the
Coma cluster. It is {\it not clear} whether the core of IC 4051 is really
{\it decoupled} from the rotation of the main body,
since the velocity curve does not drop to exactly 0 at the break radius r$_b$.
The velocity dispersion is $\sigma_{b}$~=~220 km/s at 
r$_b$ and its gradient gets much
steeper inside the core,
rising to a central value of $\sigma_o \approx 280$ km/s. Note that
the H$_3$ parameter is negative for the co-rotating core. J{\o}rgensen
et al. (1995) as well as Davies et al. (1987) 
measured the central velocity dispersion for this galaxy,
deriving values in good agreement with ours.

The Mg$_b$ index is unusually high, showing values up to 5.5~\AA\ in
the center of the galaxy. It is constant in the core region, but
decreases strongly beyond r$_b$. The asymmetry of the Mg$_b$
profile in the center of the galaxy as well as in the outer parts
is probably due to contamination from
[NI] emission lines in the interstellar medium (Goudfrooij \& Emsellem 1996).
[NI] emission can artificially increase the measured Mg$_b$ index.
The asymmetry itself arises from the fact that the stars and the gas may
have different velocities up to 100 km/s (Bertola et al. 1995). Additionally,
as shown by
Goudfrooij \& Emsellem (1996), the Mg$_2$ index is much less disturbed by the
[NI] emission lines, because of its different continuum definition.
However, the S/N and spectral resolution of our data are insufficient to
quantify this effect.
In this galaxy the mean Iron index
$<$Fe$>$ shows a steep increase towards the center
starting at the break radius, while the combined index $[$MgFe$]$
is constantly increasing towards the center of the
galaxy.  A steep decrease of the H$_\beta$ absorption index is present
outside 7''. The inspection of the H$_{\beta}$ absorption lines 
shows that they might be partially filled 
by emission like features outside this radius. As above, the S/N and
resolution of the data do not allow to quantify the effect.
Central values of Mg$_2$ are provided again by J{\o}rgensen et
al. (1995) and Davies et al. (1987) and agree with our data.\\

\section{ Photometry}

\subsection{HST observations \& data analysis}

HST images were available for both galaxies, NGC~4816
(Principal Investigator: John Lucey; Proposal ID:~5997) and IC~4051 (Principal
Investigator: James Westphal; Proposal ID:~6283). For NGC~4816 exposures
with 800~s in total are available with the F606W Filter, while for IC~4051
eight full-orbit WFPC2 exposures (20500~s in total) with the F555W
Filter ($\lambda_c~=~5957$~\AA) and two (5200~s in total) with F814W
($\lambda_c~=~7940$~\AA) were obtained. Cosmic ray
elimination and the production of a median frame from multiple
observations were carried out within ''STSDAS'' packages of the image
processing system IRAF \footnote {IRAF is distributed by the National
Optical Astronomy Observatories, which are operated by the Association
of Universities for Research in Astronomy, Inc., under cooperative
agreement with the National Science Foundation of the USA.} (Tody
1993). \\ The subsequent analysis was done within MIDAS using the data
of the Planetary Camera (PC), which has a scaling of
0.04555''/pixel. The sky values were determined from the Wide Field
Camera (WFC) images in areas which were at least 55'' away from the
galaxy's center.  After sky subtraction the foreground stars and
galaxies were either replaced with values from the opposite
side of the galaxy or totally removed and excluded from the analysis.
The photometric zero points were derived and converted to the
Johnson-Cousins system following the procedure described in
Holtzman et al. (1995).  We added a constant shift of 0.115 mag to
correct for infinite aperture and determined the I(F814W), V(F606W)and
V(F555W) band
zero points to be 21.67 mag, 23.33 mag and 22.57, respectively.

The deviations of the isophotes from ellipses were studied by means of
Fourier series expansions (Bender \& M\"ollenhoff, 1987).  The
resulting fourth cosine coefficients $a_4$ indicates boxy (negative
values) or disky isophotes (positive values). Additionally, we
estimated the brightness profile of the existing central disks with
the decomposition method developed by Scorza \& Bender (1995). The
principal idea is to consider galaxies with a positive $a_4$ parameter
to be a composition of an ideal elliptical bulge with $a_4 = 0$ and a thin
disk. The model disk is described by an exponential intensity profile
and parameterized by the scale length r$_h$, the central surface
brightness $I_0$, the position angle P.A. and the inclination
angle\footnote{where $i$ = 0$^{\circ}$ means face on, $i$ = 90$^{\circ}$
edge on.} $i$. These free parameters are varied in an iterative
process until a bulge having perfectly elliptical isophotes
($a_4 \approx$ 0) remains. We also determined the ratio of the disk's and
bulge's
luminosity (D/B = L$_D$/(L$_G$ - L$_D$))
for the whole galaxy and
inside the kinematically detected break radius r$_b$.
For the distance module of the Coma cluster we used $(m - M) = 35.7$ mag
and the V and I absolute magnitude of the Sun are
$M_{\odot,V} = 4.84$ mag and $M_{\odot,I} = 4.03$ mag. 

For IC 4051 we also derived the V-I colour
and its gradient. Since the PSF of the HST is different for the two
filters and varies with the position on the CCD, we convolved each
frame with the PSF\footnote{ The PSFs where obtained with the TINYTIM
routine.} of the other filter. Finally we get the same resulting PSF
for both frames.

\begin{table*}
\begin{center} \parbox{12cm}{ 
\begin{tabular}{|c|c|c|c|c|c|c|c|c|}
\hline
Galaxy & Filter & m$_{d}$ & r$_h$ & i & PA$_{d}$ & L$_{d}$ & (D/B)$_{tot}$ & D/B(r $\leq$ r$_b$) \\
 & & [mag] & [''] &  &  & [$10^9 L_{\odot}$] & &\\ 
\hline 
\hline
NGC 4816 & V &
17.79 & 0.94 & 64$^{\circ}$ & 100$^{\circ}$& 1.26 & 1.0 \% & 6.9 \% \\
\hline 
\hline
IC 4051 & V & 19.12 & 0.74 & 65$^{\circ}$ & 90$^{\circ}$ & 0.37 & 0.4 \% & 2.5 \% \\
IC 4051 & I & 17.53 & 0.90 & 65$^{\circ}$ & 90$^{\circ}$ & 0.79 & 0.5 \% & 2.8 \% \\
\hline
\end{tabular}
\hspace*{5cm}\caption{Photometric parameters for the central disks of NGC 4816 and IC 4051}
\label{tdiskprop}
}
\end{center}
\end{table*}

The values of the total magnitude m$_T$, the surface brightness SB and
the colour have been corrected for galactic extinction an {\it K} --
corrected. We have used the B band galactic extinction determined by
Burstein et al. (1987) for both NGC~4816 ({\it A}$_B$ = 0.03 mag) and
IC~4051 ({\it A}$_B$ = 0.05 mag). For the filters we used, we
assumed {\it A}$_{\lambda}$ = k$_{\lambda}${\it A}$_B$ and following
Seaton (1979) we determined the values k$_V$ = 0.79, k$_R$ = 0.58 and
k$_I$ = 0.48. For the {\it K} -- correction we used
values provided by Rocca-Volmerange \& Guiderdoni (1988) 
({\it K}$_V$ = 0.041 mag, {\it K}$_R$ = 0.028 mag,
{\it K}$_I$ = 0.0 mag). Finally, we corrected the surface brightness
for cosmological dimming.\\

\subsection{MDM observations \& data reduction}

We also obtained Kron-Cousins R band CCD surface photometry for NGC 4816
at the 1.3 m telescope of the MDM observatory
in March 1995.  We used a Loral 2048 x 2048 chips with a scaling of
0.508'' per pixel and observed NGC 4816 for 500s with a PSF of 2''.
Sky subtraction, cosmic ray elimination and the removal of foreground
stars were performed with MIDAS packages. We carried out the same
isophote analysis and surface brightness fit as for the HST data.  The
zero point was estimated by using V band aperture photometry from
Burstein et al.(1987) and using a typical colour for elliptical
galaxies V-R~=~0.65 mag (Poulain \& Nieto 1994). Total magnitudes m$_T$
and half luminosity radii r$_e$ were derived using the algorithm of
Saglia et al. (1997a), which fits the surface brightness profiles as a
sum of seeing convolved r$^{1/4}$ and exponential components.

\subsection{Isophotal analysis and photometric profiles}

\noindent{\bf NGC 4816:}

\begin{figure*}  
\psfig{figure=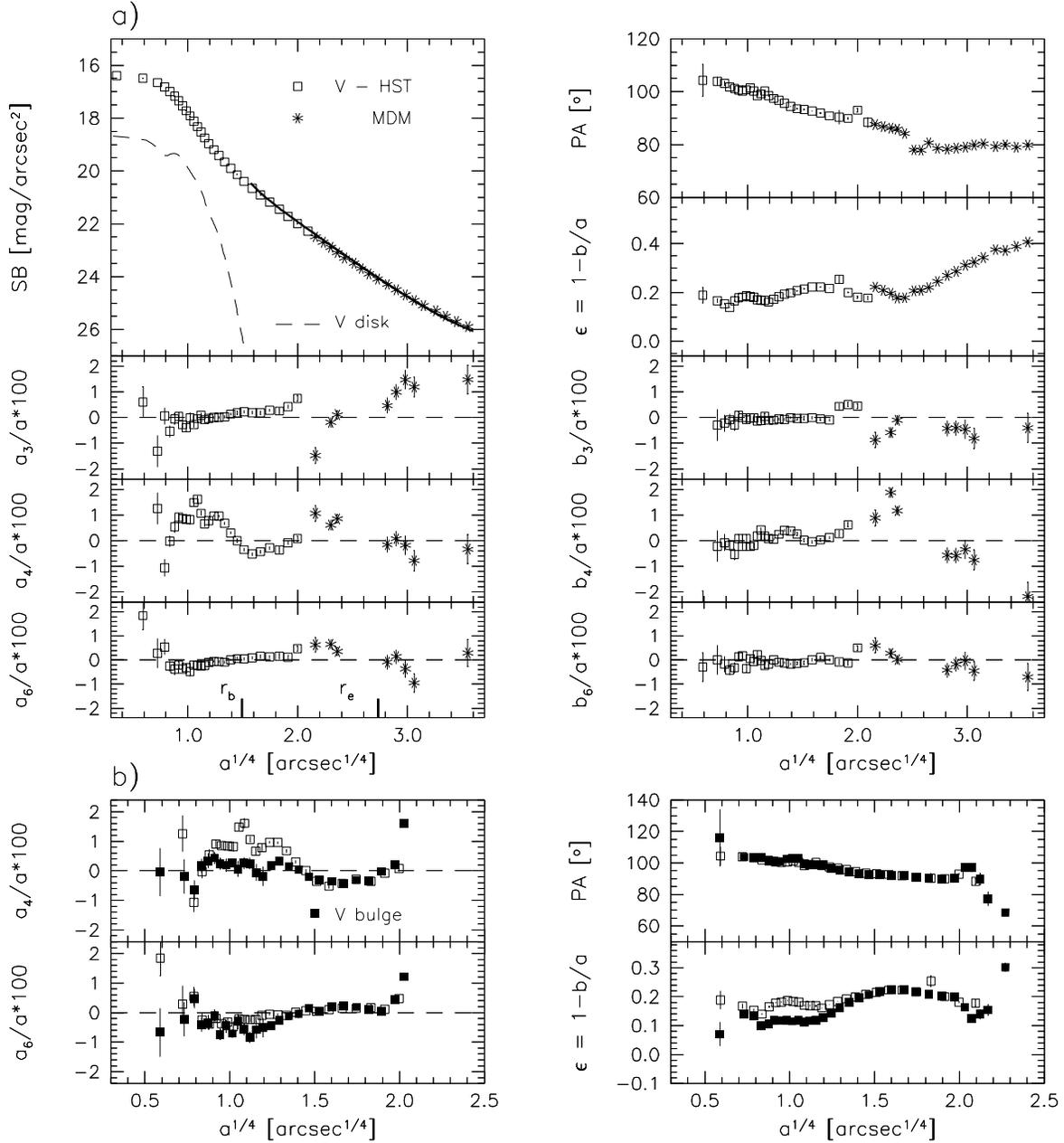,height=18cm}
\caption{\label{phot4816}
{\bf (a)} Results of the isophote analysis of NGC 4816 from the HST V band
data (open symbols) are plotted versus the major axis a. "Colour shifted"
MDM R band data are appended for the outer part of the galaxy (a $>$ 20'').
The best fit of the MDM surface brightness profile - a sum of an
r$^{1/4}$ law and an exponential component - is overplotted (fat
solid line). The fit is only shown in the outer region, since the
groundbased MDM profile starts to deviate from the HST profile inside 6''
due to different PSFs. Left panels (top to bottom): (1) Surface brightness
profile SB of the whole galaxy and the central disk only (dashed line),
(2-4) Fourier coefficients $a_3$/a, $a_4$/a, $a_6$/a. Right panels (top to
bottom): (1) Position angle PA, (2) ellipticity $\epsilon$, (3-5) Fourier
coefficients $b_3$/a, $b_4$/a $b_6$/a. {\bf (b)} Results of the
decomposition for the HST data (open symbols - whole galaxy; filled symbols
- bulge only) plotted versus the major axis a. Left panels (top to
bottom):(1-2) Fourier coefficients $a_4$/a, $a_6$/a. Right panels (top to
bottom): (1) Position angle PA, (2) ellipticity $\epsilon$. }
\end{figure*}

\noindent Fig.~\ref{phot4816}a shows the surface brightness profiles as well
as the results of the isophotal analysis for the HST V band.
We appended the "colour shifted" MDM R band data in the outer part of the
galaxy ($>$ 20''). The best fit of the MDM surface brightness profile is 
given by a sum of an r$^{1/4}$ law and an exponential 
component and is also shown in the figure.
Inside the kinematically detected break
radius (4''), the surface brightness from the HST data shows the
exponential profile typical for a disk. Additionally the $a_4$ parameter
is positive inside the core region.
Outside the core the amplitudes of $a_4$ parameter become slightly
negative. The
$a_6$ parameter is slightly negative inside the core, while
the amplitudes for the other Fourier coefficients are small
and the mean ellipticity $\epsilon$ = 0.2. 
We also show the surface brightness profile of the
best model disk from the decomposition. For the remaining bulge we
plot the $a_4$ and $a_6$ parameter, the P.A. and $\epsilon$ in
Fig.~\ref{phot4816}b together with the data values for the whole
galaxy.  The corrected photometric parameter for the central disk we
determined from the HST data are presented in Table~\ref{tdiskprop}.
Additionally m$_T$ and r$_e$ derived from the MDM data for the whole
galaxy and those available from the literature are listed
in Table~\ref{tlitphot}. The discrepancy between the r$_e$ we measured and
the literature values is due to different maximal radial extension of the
profiles. Our surface brightness profile
extends to $\approx$ 170'', while
the profiles of Andreon et al. (1996, 1997), Lucey et al. (1991) and Burstein 
et al. (1987) only reach to $\approx$~30'', 50'' and 40'', respectively.
Additionally only a pure r$^{1/4}$ profile was fitted, taking no account
of a possible exponential component.\\

{\small
\begin{table}
\begin{center}
\begin{tabular}{|c|c|c|c|c|c|c|c|c|c|}
\hline
Galaxy & Filter & m$_T$ & r$_e$ & Source \\
 & & [mag] & [''] &  \\ 
\hline 
\hline
NGC 4816 & R & 11.61 & 55.75 & this work\\
\hline
NGC 4816 & R & 11.62 & 27.54 & Andreon et al. (1996)\\
NGC 4816 & V & 12.66 & 25.12 & Andreon et al. (1997)\\
NGC 4816 & V & 12.90 & 17.14 & Lucey et al. (1991)\\
NGC 4816 & V & 12.82 & 21.20 & Burstein et al.(1987)\\
NGC 4816 & B & 13.79 & 21.20 & Burstein et al.(1987)\\
\hline 
\hline
IC 4051 & R & 12.46 & 15.56 & Saglia et al. (1997b)\\
\hline
IC 4051 & R & 12.77 & 10.72 & Andreon et al. (1996)\\
IC 4051 & r & 12.82 & 19.50 & J{\o}rgensen et al. (1992)\\
IC 4051 & V & 13.10 & 18.66 & Lucey et al. (1991)\\
IC 4051 & V & 13.04 & 20.30 & Burstein et al.(1987)\\
IC 4051 & B & 13.92 & 22.39 & J{\o}rgensen et al. (1992)\\
IC 4051 & B & 14.01 & 20.30 & Burstein et al.(1987)\\
\hline
\end{tabular}
\caption{Photometric parameters from literature (and own photometry) for NGC 4816 and IC 4051} 
\label{tlitphot}
\end{center}
\end{table}
}

\noindent {\bf IC 4051:}

\begin{figure*}  
\psfig{figure=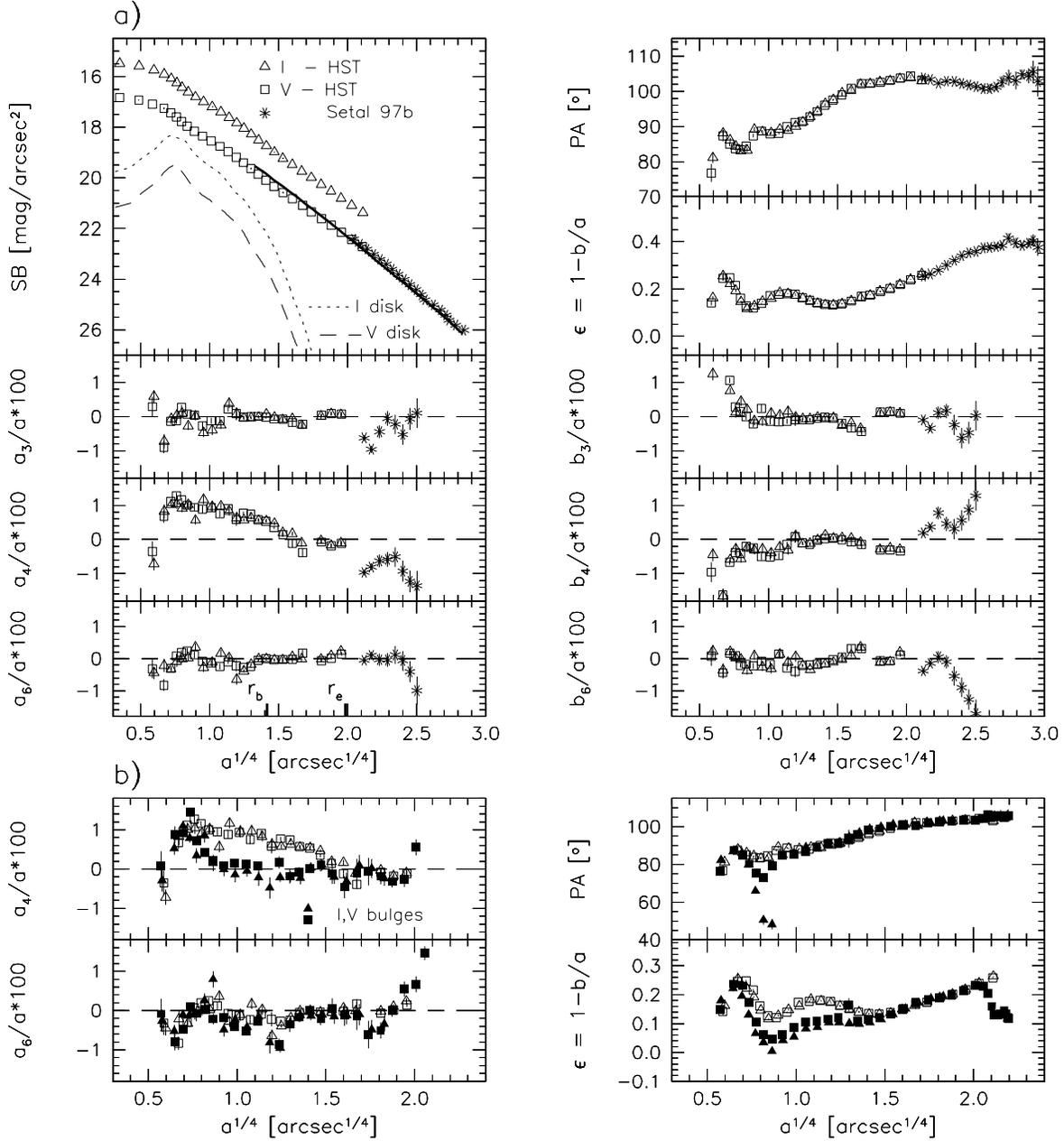,height=18cm}
\caption{\label{phot4051}
{\bf (a)} Results of the isophote analysis of IC 4051
from the HST V and I band
data (open symbols) are plotted versus the major axis~a. "Colour shifted"
R band data from Saglia et al. (1997b) (stars) are appended for the outer part 
of the galaxy (a $>$ 20''). The best fit of the R band 
surface brightness profile
 - a sum of an r$^{1/4}$ law and an exponential
component - is overplotted (fat solid line). The fit is only shown in the
outer region, since the groundbased R band profile starts to deviate from the HST profile
inside 3'' due to different PSFs. 
Left panels (top to bottom): (1) Surface brightness profile SB of the 
whole galaxy and the central disk only (short and long dashes), (2-4) 
Fourier coefficients $a_3$/a, $a_4$/a, $a_6$/a. 
Right panels (top to bottom): (1) Position angle PA, (2) ellipticity
$\epsilon$, (3-5) Fourier coefficients $b_3$/a, $b_4$/a $b_6$/a. 
{\bf (b)} Results of the decomposition for the HST data 
(open symbols - whole galaxy; filled symbols - bulge only) 
plotted versus the major axis a.
Left panels (top to bottom):(1-2) Fourier coefficients $a_4$/a, $a_6$/a. 
Right panels (top to bottom): (1) Position angle PA, (2) ellipticity
$\epsilon$.
}
\end{figure*}

\begin{figure}  
\psfig{figure=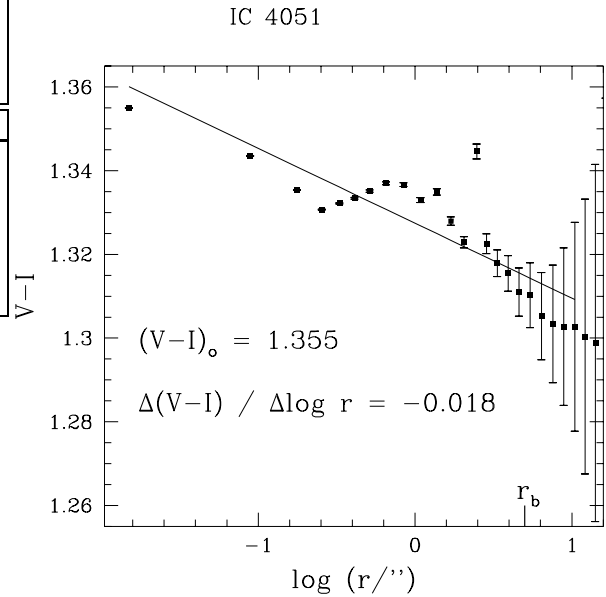,height=8cm}
\caption{\label{colour}
V - I colour profile of IC 4051 and the results of the least 
square fit for the colour gradient within 10''.}
\end{figure}

\noindent The surface brightness profiles and the results of the
isophotal analysis are shown in Fig.~\ref{phot4051}a for both, the
Johnson I and the V band exposure. We appended the "colour shifted"
R band data from Saglia et al. (1997b) in the outer part of the galaxy
($>$ 20''). The best fit of the R band surface brightness profile is given by 
a sum of an r$^{1/4}$ law and an exponentiell component 
and is also shown in the figure.
In both the I and the V band the $a_4$ parameter is
positive inside 5'', which is the size of the co-rotating decoupled
component, detected in the velocity curve. Additionally, the 
$a_6$ parameter is slightly negative inside the core, while the 
amplitudes for the other
Fourier coefficients are small and the mean ellipticity 
$\epsilon$~=~0.2. We also show the surface brightness profile of the best model
disk from the decomposition. We plot the $a_4$ and $a_6$
parameter, the P.A. and $\epsilon$ for the remaining bulge in 
Fig.~\ref{phot4051}b together with the results for the whole galaxy. The
corrected photometric parameter for the central disk we determined
from the HST data are presented in Table~\ref{tdiskprop} for both
filters. Additionally, the total magnitude m$_T$ and the effective
radius r$_e$ for the whole galaxies available from the literature are
listed Table~\ref{tlitphot}. 
The central colour (V-I)$_o$ is 1.355 mag and the V-I
colour profile is show in Fig.~\ref{colour}. We derived a mean
colour gradient $\Delta(V-I)/\Delta log(r)$
within 10'' of -0.018 mag/dex by a linear fit. 
For comparison,  J{\o}rgensen et al. (1992) found a
B-r gradient of -0.09 mag per dex in radius inside r$_e$.

\section{Results} \label{results}

\begin{figure*}  
\psfig{figure=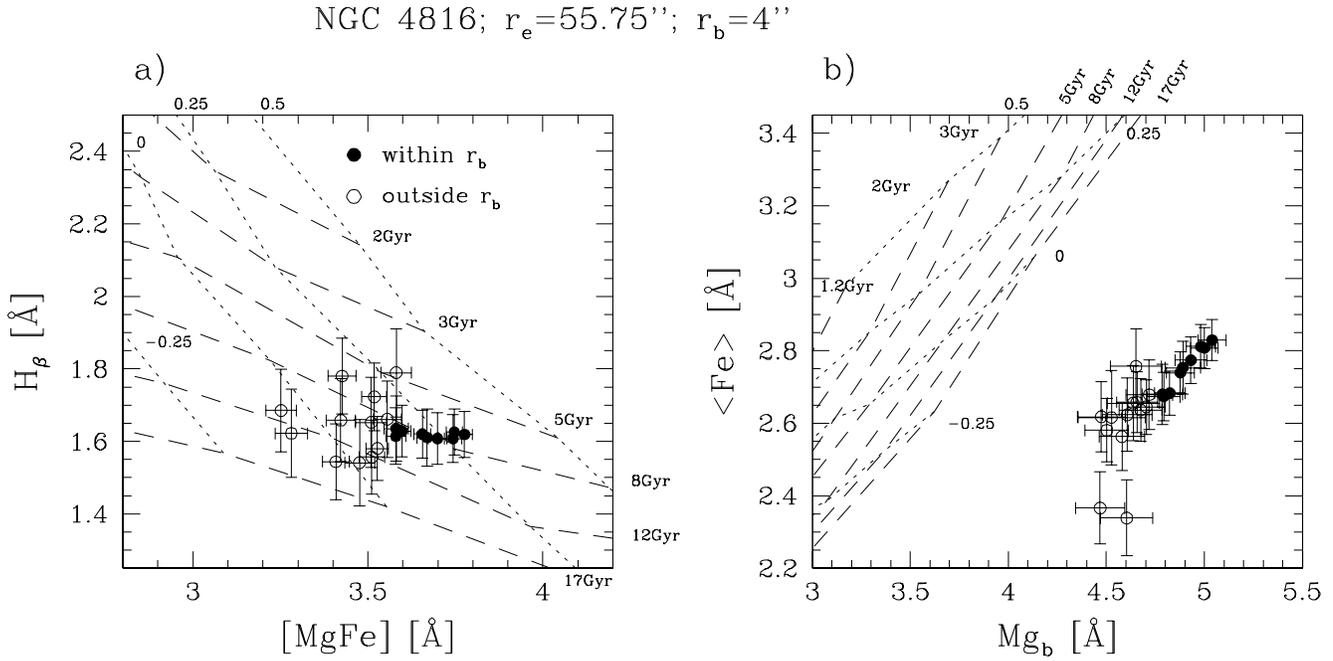,width=17.5cm}
\caption{\label{AZmodel4816}
Measured absorption line indices of NGC 4816 
plotted into the model grids of Worthey
(1994); dashed lines - constant age in Gyr; dotted lines  -
constant metallicity log(Z/Z$_{\odot}$). Filled symbols represent
measurements inside the break radius r$_b$, open symbols those for
measurements outside r$_b$.
{\bf (a)} The age indicating H$_{\beta}$ versus 
the metallicity indicating combined index $[$MgFe$]$. 
{\bf (b)} $<$Fe$>$ versus Mg$_b$ index. Obviously the models cannot
reproduce the measured element ratios.
}

\end{figure*}

The early -- type galaxy {\bf NGC 4816}
hosts a {\it decoupled} counter rotating {\it core} with a size of
r$_b$ = 4'' (Fig.~\ref{spec4816}a).
Inside this core the H$_3$ parameter, which
measures the asymmetric deviations from a gaussian velocity
distribution, is slightly positive where v $<$ 0.
This confirms
Bender et al. (1994), who find that H$_3$ becomes
systematically negative with increasing v/$<\sigma>$. Hence, the LOSVD in the
core of NGC 4816 are {\it intrinsically asymmetric}, showing an excess of
fast rotating stars.\\
The isophotes inside 4'' are clearly disky, which is derived from the
positive a$_4$ parameter from the isophotal shape analysis
(Fig.~\ref{phot4816}a).
Since the disky isophotes extend to the same radius as
the counter rotating core, detected in the velocity curves, it is
evident that this component is a {\it central stellar disk}.
Additionally, the decomposition of the surface
brightness profiles into a central exponential disk profile and a bulge
following an r$^{1/4}$ law, yields a perfect remaining bulge with
$a_4$~=~0 (Fig.~\ref{phot4816}b).
The central disk in NGC~4816 contributes only 1~\% to the total
V band light of the galaxy.
This is even less than found for the kinematically decoupled central disk 
inside NGC~5322 (Bender et al. 1994), which has D/G~=~3~\%
(Scorza \& Bender 1995). 
Inside the break radius r$_b$~=~4'' the central disk of NGC~4816 
contributes 6.9~\% to the total light.
For a thin disk one would expect a positive $a_6$
parameter from the isophote analysis (Scorza \& Bender 1995), but not
a negative one as found in our data. Thus, with the assumption of having a
{\sf thin} central disk in NGC 4816 the ``measured''
inclination is only a lower limit.\\
The slope of the Mg$_b$ line strength profile
steepens inside
the break radius r$_b$ (Fig.~\ref{spec4816}b), which points to a metal
enriched core and agrees with results found for other peculiar cores by
Bender \& Surma (1992). The
mean iron index $<$Fe$>$ and the combined index $[$MgFe$]$ do also
increase inside 8''.
In the rotation curve the core is only showing up at
4''. As it is {\sf counter} -- rotating it has to compete against the 
superposed
rotation of the bulge further out and becomes only dominant inside 4''.
In fact in (Fig.~\ref{spec4816}a) you can see that the rotation curve seems
to be influenced by the core out to 6'' or 7''.
Comparisons with stellar population
models (Worthey 1994) show that the 
{\it metallicity}~\footnote{Following Faber et al. (1995) we used the 
combined index
$[$MgFe$]$ = (Mg$_b$ $<$Fe$>$)$^{1/2}$, which represents the mean metallicity
best.}
of the core is Z~=~log~((Fe/H)/(Fe/H)$_{\odot}$)~$\approx$~0.25~dex
{\it super solar} (Fig.~\ref{AZmodel4816}a), 
which is comparable to values measured
in central regions of normal galaxies (Faber et al 1995). 
On the other hand the metallicity of the hosting galaxy
drops to values around solar metallicity
(outer most data point at $\approx$~0.5~r$_e$).
In the core as well as in the outer parts of the galaxy 
the dominating {\it stellar population is old} (8 -- 12 Gyr) 
(Fig.~\ref{AZmodel4816}a).\\
From other studies it is already known that
luminous, massive elliptical galaxies can reach a
Mg$_b$ to $<$Fe$>$ overabundance
[Mg/Fe] $\approx$ 0.4 dex super solar, while faint early -- type 
galaxies have $\approx$ solar abundances 
(Gonzales 1993; Fisher et al. 1995).
NGC 4816 shows extreme Mg$_b$ to $<$Fe$>$ {\it overabundance} of
$\approx$~0.5~dex (Fig.~\ref{AZmodel4816}b) 
with no significant abundance 
difference between the core and its surrounding, which is in
agreement with other investigations (Davies et al. 1993; Paquet, 1994).
In a recent study, Thomas et al. (1997) show that
in addition to a top heavy initial mass function (IMF), a short star
formation time scale is needed to explain this light
element enrichment in massive elliptical galaxies with an excess of
Supernovae II rate.\\

\begin{figure*}  
\psfig{figure=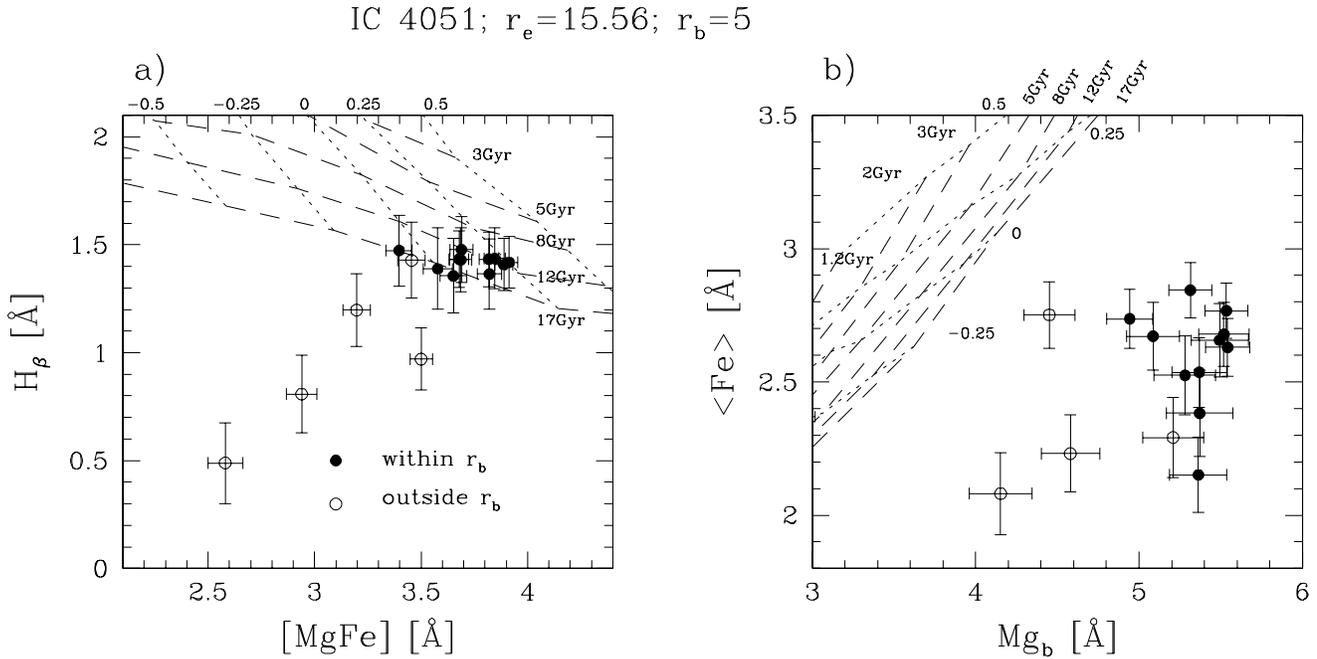,width=17.5cm}
\caption{\label{AZmodel4051}
Measured absorption line indices of IC 4051 plotted into 
the model grids of Worthey
(1994); dashed lines - constant age in Gyr; dotted lines  -
constant metallicity log(Z/Z$_{\odot}$). Filled symbols represent
measurements inside the break radius r$_b$, open symbols those for
measurements outside r$_b$.
{\bf (a)} The age indicating H$_{\beta}$ versus 
the metallicity indicating combined index $[$MgFe$]$. Outside r$_b$ the
H$_{\beta}$ indices are emission contaminated and hence not reliable for age
determination. 
{\bf (b)} $<$Fe$>$ versus Mg$_b$ index.  Obviously the models cannot
reproduce the measured element ratios.
}
\end{figure*}

For {\bf IC 4051} there is evidence for a co-rotating {\it peculiar core} 
with a radial size of r$_b$ = 5'' (Fig.~\ref{spec4051}a).
In contrast to the core NGC 4816 it is not clear whether
this central component in IC 4051 is really decoupled.
From the presented kinematic data (Fig.~\ref{spec4051}a), we cannot exclude 
that IC 4051 is simply a disky elliptical or S0 galaxy being a superposition 
of a hot (rotating) bulge and a cold, fast rotating disk 
(Scorza \& Bender 1995). In combination with the brightness 
profiles we derived from the HST images, we can compute the 
anisotropy parameter (Binney 1978, Kormendy 1982):
\begin{equation}
\left( \frac{v_r}{\sigma_m} \right)^\ast  =
\frac{v_r/\sigma_m}{\sqrt{\epsilon/(1-\epsilon)}} ,
\end{equation}

where $v_r$ is the rotation velocity, $\epsilon$ the ellipticity and 
$\sigma_m$
the averaged velocity dispersion. Objects with $(v/\sigma)^\ast > 0.7 $ are 
mainly rotationally flattened, while those having smaller values are 
flattened by anisotropic velocity dispersions (Davies et al. 1983).
Since the anisotropy parameter for IC 4051 is $\approx$ 0.5 inside the "core" 
and $\approx$ 0.4 at r$_e$ it is rather unlikely that it is an S0 galaxy.\\
From the isophotal analysis and the photometric decomposition 
(Fig.~\ref{phot4051}) we 
inferred that this core is a {\it central stellar disk}.
Like the disk in NGC 4816 it only
contributes 0.4~\% to the total V band light of the galaxy and
0.5~\% to the total I band light. Since the disks in ``normal'' S0 galaxies
contribute $\approx$ 4 -- 11~\% to the total V band light (Scorza \& Bender
1995) IC 4051 is probably {\it not} an S0 galaxy, but contains a decoupled
central disk.
Inside the break radius r$_b$~=~5'' the disk of IC 4051
contributes 2.4~\% and 2.7~\% in the V and I band, respectively.
The V-I colour gradient inside the IC 4051's core is
shallow and in agreement with Carollo et al. (1997) it does not indicate
any significant difference between the core and the surrounding galaxy.\\
In contrast to NGC 4816 the slope of the Mg$_b$ line profile does not increase
inside the core of IC 4051 (r $<$ 5'') but is constant 
(Fig.~\ref{spec4051}b).
This lack of steepening inside the core could be due to [NI] emission that 
can contaminate the Mg absorption feature (Goudfrooij \& Emsellem 1996).
On the other hand the gradient of the mean iron index
$<$Fe$>$ is enhanced towards the center starting at the break radius,
pointing to a {\it metal enriched} core as found in NGC 4816.\\
Comparisons with stellar population
models (Fig.~\ref{AZmodel4051}a) show that the {\it metallicity} drops from
{\it super solar} inside the core ($\approx$~0.25~dex) to
-0.5 dex in the outer part of the galaxy (outer most data point at 
$\approx$~1~r$_e$). This metallicity difference of 0.75~dex
between the central and the embedding galaxy
is unusually high.
The stars in the core are {\it old} (12 --17 Gyr), while an
age estimate in the outer parts is not possible since the
H$_\beta$ absorption lines are probably contaminated by emission like
features in the outer parts of the galaxy.
As in NGC~4816 the dominating stellar population is {\it highly} $[$Mg/Fe$]$
{\it overabundant}. 


\section{Discussion}

NGC 4816 contains a kinematically decoupled central disk, which is dominated
by an old, metals enriched, highly $[$Mg/Fe$]$
overabundant stellar population and is only contributing $\approx$ 1~\% to
the total light of the galaxy. 
For the formation scenario of this disk we favour 
{\it dissipational major merging} for the following reasons:\\
\noindent $\bullet$ A {\it merging} event is necessary to account
for the kinematic decoupling between core and bulge 
(Barnes 1992b).\\
\noindent $\bullet$ {\it Dissipational} merging is required in contrast to 
dissipationless processes because: (1) In a
dissipational merger the gas can get to the center and form a 
{\it flattened, disk -- like} core by dissipation and cooling (Barnes
\& Hernquist 1996). Flattened systems like the core of NGC 4816
cannot be built through dissipationless processes.
(2) While dissipationless merger tend to weaken metallicity
gradients (White 1980), 
only dissipative processes can built up these gradients. 
Previously enriched gas settles in the core and may trigger the
formation of {\it metal enriched} stars (Bender \& Surma 1992) .
The enhancement of the stellar metallicity
we find in the central disk of NGC 4816 require gas dissipation.\\
\noindent $\bullet$ The accretion of low-mass gas 
rich galaxies can be ruled out since they are
generally too metal poor (Bender \& Surma 1992). 
To obtain the high core metallicity measured in
NGC~4816 a {\it major} merging of massive, star dominated, but gas-rich
pro-genitors is needed.\\
\noindent $\bullet$ Finally, merging or interacting galaxies experience
a burstlike star formation, that may produce flatter IMFs (Wright et al.
1988; Silk 1992) and hence could help to
produce the $[$Mg/Fe$]$ {\it overabundance} we measure in the stellar
population of NGC~4816.\\

Since the stellar population of the central disk is likely
old, this merging events must have
occured at high redshift in the early phase of formation. However, the age 
indicating H$_{\beta}$ line
strength is strongly sensitive to H$_{\beta}$ emission from younger 
populations. Hence, the estimated age is only an upper limit to the true age of the
stars in the central disk. If the true age was significantly younger,
the central disk in NGC 4816 could also have formed in late, gas-rich 
mergers, which are e.g. observed in ultra-luminous IRAS galaxies
(Solomon et al. 1992, Prestwich et al. 1994). However, late mergers may have
problems in producing the light element overabundances.\\

The central disk in IC 4051 is also dominated by an old, metal rich, 
highly $[$Mg/Fe$]$
overabundant stellar population and contributes only $\approx$ 1 \% to
the total light of the galaxy. In contrast to NGC 4816 the interpretation of 
the data with respect to the formation of the galaxy and its peculiar core 
is less evident.
Whether the disk in IC 4051 is {\it really} kinematically {\it decoupled}
and hence whether this galaxy is indeed a major merger remnant
remains an open question, which can not be answered with our data.\\

\begin{table*}
\begin{center}
\begin{tabular}{|c|c|c|c|c|l|}
\hline
Galaxy & vhel (km/s)/  & scaling & r$_b$ & r$_e$ & References	\\
  & Distance (Mpc) & & & & \\  
\hline 
\hline
 I 4051	&  7000 / 140 & 1'' = 678 pc & 5'' = 3.4 kpc & 16'' = 10.5 kpc & this paper\\	
 N 4816 & 7000 / 140 & 1'' = 678 pc & 4''= 2.7 kpc & 55.75'' = 37.8 kpc & this paper\\	
\hline
 N 4365$^{\ast}$ & 1153 / 23 & 1'' = 112 pc & 8'' = 0.9 kpc & 64'' = 6.4 kpc & BS92, BSG94, B88\\
 N 4406$^{\ast}$ & 1153 / 23 & 1'' = 112 pc & 6'' = 0.7 kpc & 199'' = 22.3 kpc & BSG94, BS92, FIH89, B88,\\
 N 4472$^{\ast}$ & 1153 / 23 & 1'' = 112 pc & 6'' = 0.7 kpc & 104'' = 11.65 kpc & Setal93, BSG94, DB88\\
 N 4494	& 1172 / 23 & 1'' = 114 pc & 7'' = 0.8 kpc & 60'' = 6.8 kpc & BSG94, BS92, B88, JS88\\
 I 1459  & 1647 / 33  & 1'' = 160 pc & 10'' = 1.7 kpc & 35'' = 5.8 kpc & FI88, FIH89\\
 N 3608	& 1126 / 23 & 1'' = 109 pc & 10'' = 1.1 kpc & 34'' = 3.7 kpc & JS88\\
 N 5322	& 1963 / 39 & 1'' = 190 pc & 10'' = 1.9 kpc & 34'' = 6.5 kpc & BSG94, B88\\
 N 5813	& 1674 / 33 & 1'' = 162 pc & 10'' = 1.6 kpc & 57'' = 9.2 kpc & EEC82\\
 N 7626	& 3493 / 70 & 1'' = 339 pc & 8'' = 2.7 kpc & 39'' = 13.2 kpc & JS88, BC93\\
 N 1427	& 1411 / 28 & 1'' = 137 pc & 8'' = 1.1 kpc & 33'' = 4.5 kpc & D'Oetat95\\
 N 1439	& 1585 / 32 & 1'' = 154 pc & 10'' = 1.5 kpc &  41'' = 6.3 kpc & FIH89\\
 N 1700	& 4082 / 82 & 1'' = 396 pc & 5'' = 2.0 kpc & 14'' = 5.5 kpc & BSG94, FIH89, SSC96\\
 N 5982	& 2787 / 56 & 1'' = 270 pc & 7'' = 1.9 kpc & 25'' = 6.8 kpc & WBM88\\
 N 7796	& 3252 / 65 & 1'' = 315 pc & 4'' = 1.3 kpc & 27'' = 8.5 kpc & Betal94\\
 N 6851 & 3051 / 61 & 1'' = 296 pc & 2.5'' = 0.7 kpc & 14.5'' = 4.3 kpc & TSZ92\\
\hline
\end{tabular}
\caption{Galaxies with kinematically decoupled cores mentioned in the literature.
We used H$_o$ = 50 km / (s Mpc) to the determine the distances. At the 
``break radius'' r$_b$ the decoupled core starts to dominate the rotation
curve. Galaxies with an $^\ast$ belong to the Virgo cluster.
}\label{tlit}
\end{center}
\end{table*}

\begin{figure}  
\psfig{figure=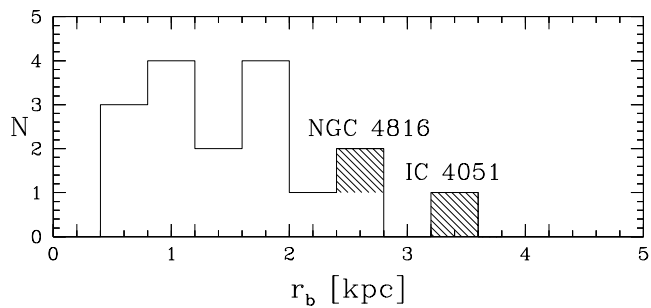,width=8.5cm}
\caption{\label{hist}
Distribution of the break radii r$_b$ of all known kinematically distinct
cores. The positions of IC 4051 and NGC 4816 are indicated by crosses.}
\end{figure}

Table~\ref{tlit} contains the 17 known early -- type galaxies with
clear evidence for a kinematically decoupled core.
Most of the other cores have sizes smaller than 2 kpc (see the histogram
Fig.~\ref{hist}).
With a break radius of r$_b$ = 2.7 kpc NGC 4816
hosts one of the largest kinematically decoupled core ever detected in
an early -- type  galaxy. Only the decoupled central of NGC~7626, the
dominating
galaxy of the Pegasus cluster (Jedrzejewski\& Schechter 1988;
Balcells \& Carter 1993) has a similar size.
The peculiar core in IC 4051 (r$_b$ = 3.4 kpc) is even bigger.
To investigate whether the large cores of NGC~4816 and IC~4051 are
systematically different to other, smaller kinematically decouples cores, we
compared the properties of their corresponding central disks. HST
images are available for 13 of the 15 galaxies that are not in Coma and are
listed in Table~\ref{tlit}. We
applied the D/B decomposition of Scorza \& Bender (1995)
to derive the properties of
the central disks, that are present in 8 of these galaxies. Since 4 of these disks
were contaminated with dust we are left with reliable disk properties for
NGC~1427, NGC~4365, NGC~5322 and NGC~5982.
Plotting the central surface brightness SB$_{\circ}$ versus the scale length
r$_h$ shows that the disks of NGC~4816 and IC~4051 simply represent the
``faint'', ``large'' end of a sequence. We therefore speculate that
the large cores
of NGC~4816 and IC~4051 had formation histories similar to the ones of the
smaller, already known decoupled cores. Furthermore, the parameters r$_h$
and SB$_{\circ}$ derived for the central disks of galaxies with cores are
similar to the ones derived for 
``disky ellipticals'' (Scorza \& Bender 1995).\\
Excluding Coma's two
dominant galaxies (NGC~4874 \& NGC~4889) and the central galaxy (NGC~4839)
of the Coma substructure in the S/W (Colless \& Dunn 1995),
IC~4051 is the third brightest galaxy in the central square degree of
the cluster (see J{\o}rgensen \& Franx 1994). NGC~4816 is even 0.75 mag
brighter than IC~4051 in the R band and hence,
both galaxies have luminosities above L$_*$.
While NGC~4816 lies 49.0' S/W of the clusters center
and only 21.8'
apart from NGC~4839, IC~4051 lies only 15.5' N/E of the clusters center.
Whereas NGC 4816 is the largest host galaxy (r$_e$ = 37.8 kpc) of a distinct 
core, the size of IC 4051 is comparable the sizes of the host galaxie listed
in Table~\ref{tlit}.
However, like the other galaxies with kinematically decoupled cores
{\it both} galaxies do not deviate from the Fundamental
Plane or the Mg -- $\sigma$ relation. Our total sample of
35 Coma galaxies is complete for galaxies brighter than
M$_B$ = -21.6 mag (14 galaxies) and is 1/3 complete
in the magnitude range -21.6 mag $< M_B <$ -20.2 mag (21 galaxies). 
NGC 4816 and IC 4051 both belong to the complete sample. Similarly, in the
Virgo cluster there are only two galaxies (NGC 4406 and NGC 4472) that
host a decoupled core and are brighter than -21.6 mag. The core of NGC~4406
contributes $\approx$~20~\% to the the V band light inside the kinematically
break radius r$_b$ (Surma 1992), 
which is consistent with the central disks in the Coma galaxies.
In the magnitude
range -21.6 mag $< M_B <$ -20.2 mag one galaxy with a decoupled core 
is known
in the Virgo cluster (NGC 4365), but none found in the Coma cluster
yet. In their detailed discussion of NGC 4365's
stellar population and kinematic
Surma \& Bender (1995) conclude that its core also must have built through
dissipational major merging processes. This core also shows disk -- like
structure, its mean stellar population has a metallicity which is
$\approx$ 0.4 dex above the Sun and an age of $\approx$~7~Gyr. Inside the
kinematically break radius r$_b$ it contributes $\approx$ 20 \% to the V band 
light.
Note that the break radius r$_b$ of the ``Virgo cores'' would
compare to $<$ 1.3'' at Coma's distance. Thus with our observational
setups (spatial resolution $\approx$ 1.5'') we missed all possibly
existing cores with comparable size to the Virgo decoupled cores.  On the
other hand there is no core detected in the Virgo cluster that has
a similar size like those found in the Coma cluster galaxies 
NGC 4816 and IC 4051. 
Based on our data, we cannot derive any statistically significant difference 
between the Coma and
the Virgo cluster with respect to the fraction of galaxies with
kinematically decoupled/peculiar cores. 

\begin{acknowledgements} The authors thank the staff of the Calar Alto
and MDM observatory for their effective support. We gratefully
acknowledge the helpful discussions with
Paola Belloni and Laura Greggio.  We also want to thank the 
referee Marcella Carollo for helpful comments.
This work was supported by the Deutsche Forschungsgemeinschaft via 
project Be 1091/6.
\end{acknowledgements}

\section{References:}

Andreon S., Davoust E., Michard R., Nieto J.-L.,\\ 
\hspace*{0.5cm} Poulain P., 1996, A\&AS 116, 429\\
Andreon S., Davoust E., Poulain P., 1997, A\&AS 126, 67\\
Balcells M., 1991, A\&A 249, L9\\
Balcells M., Carter D. 1993, A\&A 279, 376 (BC93)\\
Balcells M., Quinn P.J., 1990, ApJ 361, 381\\
Barnes J., 1992a, ApJ 393, 484\\
Barnes J. 1992b, ApJ 471, 115\\
Barnes J., Hernquist L., 1996, ApJ 471, 115\\
Bender R., 1988, A\&A 202, L5 (B88)\\
Bender R., 1990, A\&A 229, 441\\
Bender R., M\"ollenhoff C., 1987, A\&A 177, 71\\
Bender R.,  Surma P., 1992, A\&A 258, 250 (BS92)\\
Bender R., Saglia R.P., Gerhard O., 1994, MNRAS 269, 785\\
\hspace*{0.5cm}(BSG94)\\
Bertin G., Bertola F., Buson L.M., et al., 1994,A\&A 292, 381\\
\hspace*{0.5cm}(Betal94)\\
Bertola F., Cinzano P., Corsini E.M., Rix H.-W.,\\ 
\hspace*{0.5cm}Zeilinger W.W., 1995, ApJ 448, L13\\
Binney J., 1978, MNRAS 183, 779\\
Burstein D., Davies R.L., Dressler A., et al., 1987, ApJS 64,\\
\hspace*{0.5cm}601\\
Carollo M., Franx M., Illingworth G.D., Forbes D.A., 1997,\\
\hspace*{0.5cm}ApJ 481, 710\\
Colless M., Dunn A., 1996, ApJ 458, 435\\
Davies R.L., Efstathiou G., Fall S.M., Illingworth G.D.,\\
\hspace*{0.5cm}Schechter P.L., 1983, ApJ 266, 41\\
Davies R.L., Burstein D., Dressler A., et al., 1987,\\
\hspace*{0.5cm}ApJS 64, 581\\
Davies R.L., Sadler E.M., Peletier R.F., 1993, MNRAS 262,\\
\hspace*{0.5cm}650\\
D'Onofrio M., Zaggia S.R., Longo G., Caon N., Capaccioli M.,\\
\hspace*{0.5cm}1995, A\&A 296, 319 (D'Oetal95)\\
Efstathiou G., Ellis R. S., Carter D., 1982, MNRAS 201, 975\\
\hspace*{0.5cm}(EEC82)\\
Faber S.M., Friel E.D., Burstein D., Gaskell C.M., 1985,\\
\hspace*{0.5cm}ApJS 57, 711\\
Faber S.M., Trager S., Gonzalez J.J., Worthey G., 1995, IAU\\
\hspace*{0.5cm}Symp. 164 ({\it Stellar Populations}),
ed. P.C. van der Kruit\\
\hspace*{0.5cm}\& G. Gilmore, Kluwer Dordrecht\\
Faber S.M., Tremaine S., Ajhar E., et al., 1997, AJ (in press)\\
Franx M., Illingworth G.D., 1988, ApJ 327, L55 (FI88)\\
Franx M. Illingworth G.D., Heckmann T., 1989, ApJ 344, 613\\
\hspace*{0.5cm}(FIH89)\\
Fisher D., Franx M., Illingworth G.D., 1995, ApJ 448, 119\\
Gonzales J.J., 1993, PhD thesis, University of California,\\
\hspace*{0.5cm}Santa Cruz\\
Goudfrooij P., Emsellem E., 1996, A\&A 306, L45\\
Hernquist L., 1993, ApJ 409, 548\\
Heyl J.S., Hernquist L., Spergel D.N., 1994, ApJ 427, 165\\
Holtzman J.A., Burrows C.J., Casertano S., et al. 1995\\
\hspace*{0.5cm}PASP 107, 1065\\  
Jaffe W., Ford H.C., O'Connell R.W., van den Bosch F.C.,\\
\hspace*{0.5cm}Ferrarese L., 1994, AJ 108, 1567\\ 
Jedrzejewski R., Schechter P.L., 1988, ApJ 330, L87 (JS88)\\
J{\o}rgensen I., Franx M., 1994, ApJ 433, 553\\
J{\o}rgensen I., Franx M., Kj{\ae}rgaard P., 1992, A\&AS 95, 489\\
J{\o}rgensen I., Franx M., Kj{\ae}rgaard P., 1995, MNRAS 276,\\
\hspace*{0.5cm}1341\\
Kormendy J., 1982, in {\it Morphology and Dynamics of Galaxies}, \\
\hspace*{0.5cm}eds. Martinet, L., Mayor, M., Saas Fee\\
Kormendy J., 1984, ApJ 287, 577\\
Kormendy J., Sanders D.B., 1992, ApJ 390, L53\\
Lauer T.R., Ajhar E.A., Byun Y.-I., et al., 1995, AJ 110, 2622\\
Lucey J.R., Guzm\'{a}n R., Carter D., Terlevich R.J., 1991,\\
\hspace*{0.5cm}MNRAS 253, 584\\
Mehlert D., 1998, PhD thesis, Universit\"at M\"unchen\\ 
Nieto J.-L., Bender R., Arnaud J., Surma P., 1991, A\&A\\
\hspace*{0.5cm}244, L25\\
Paquet A., 1994, PhD thesis, University of Heidelberg\\
Poulain P., Nieto J.-L., 1994, A\&AS 103, 573\\
Prestwich A.H., Joseph R.D., Wright G.S., 1994, ApJ 422, 72\\
Rix H.-W., White S.D.M., 1992, MNRAS 254, 389\\ 
Rocca-Volmerange B., Guiderdoni B., 1988, A\&AS 75, 93\\ 
Saglia R.P., Bertin G., Bertola F., et al., 1993, ApJ 403, 567\\
\hspace*{0.5cm}(Setal93)\\
Saglia R.P., Bertschinger G., Baggley G., et al., 1997a,\\
\hspace*{0.5cm}ApJS 109, 79 \\
Saglia R.P., Burstein D., Bertschinger G., etal., 1997b,\\
\hspace*{0.5cm}MNRAS (in press; Setal97b)\\
Schweizer F., 1990, in `Dynamics and Interactions of \\
\hspace*{0.5cm}Galaxies', ed. R. Wielen, Springer -- Verlag, p.60\\
Scorza C., Bender R., 1995, A\&A 293, 20\\
Seaton M.J., 1979, MNRAS 187, 785\\
Silk J., 1992, , IAU Symp. 149 ({\it The Stellar Populations of}\\
\hspace*{0.5cm}{\it Galaxies}), ed. B. Barbuy, Kluwer Dordrecht\\
Solomon P.M., Downes D., Radford S.J.E., 1992, ApJ 387, L55\\
Statler T.S., Smecker-Hane T., Cecil, G.N, 1996, AJ 111,1512\\ 
\hspace*{0.5cm}(SSC96)\\
Surma P., 1992, PhD thesis, Universit\"at Heidelberg\\ 
Surma P.,  Bender R., 1995, A\&A 298, 405\\ 
Thomas D., Greggio L., Bender R., 1997, MNRAS (in press)\\
Tody D., 1993, Astronomical Data Analysis Software and \\
\hspace*{0.5cm}Systems II, A.S.P. Conference Series 52, eds. R.J. Hanisch,\\
\hspace*{0.5cm}R.J.V. Brissenden, J. Barnes\\
Toniazzo T., Stiavelli M., Zeilinger W.W., 1992, A\&A 259, 39\\ 
\hspace*{0.5cm}(TSZ92)\\
van der Marel R.P., Franx M., 1993, ApJ 407, 525 \\
Wagner S.J., Bender R., M\"ollenhoff C., 1988, A\&A 195, L5\\
\hspace*{0.5cm}(WBM88)\\
White S.D.M., 1980, MNRAS 191, 1\\
Worthey G., 1992, PhD thesis, University of California, Lick\\
\hspace*{0.5cm}Observatory\\
Worthey G., 1994, ApJS 95, 105\\
Wright G.S., Joseph R.D., Robertson N.A., James P.A.,\\
\hspace*{0.5cm}Meikle W.P.S, 1988, MNRAS 233, 1

\end{document}